\documentclass[12pt,a4paper,final]{iopart}

\usepackage{iopams}  
\usepackage{graphicx}
\usepackage[breaklinks=true,colorlinks=true,linkcolor=blue,urlcolor=blue,citecolor=blue]{hyperref}
\begin{document}
\title{Fluorine absorption on single and bilayer graphene: Role of sublattice and layer decoupling}

\author{Hern\'an Santos$^{1,2}$ and Luc Henrard$^2$}
\address{$^1$ Departamento de F\'{\i}sica Fundamental, Universidad Nacional de Educaci\'on a Distancia, Apartado 60141, E-28040 Madrid, Spain}
\address{$^2$ Research Center in Physics of Matter and Radiation (PMR), University of Namur, Rue de Bruxelles 61, B-5000 Namur, Belgium}
\eads{\mailto{hernan.santos@fisfun.uned.es}, \mailto{ luc.henrard@unamur.be}}

\begin{abstract}

The fluorination of mono- and bi-layer graphene have been studied by
means of ab-initio DFT calculations. The stability of CF$_x$ systems
are found to depend on both the F coverage and on the position of the F 
atoms regarding the C sublattices. When F atoms is chemisorbed to C atoms
belonging to the same sublattice, low coverage is preferred. Otherwise,
large F coverable is more stable (up to C$_4$F). The difference of charge
distribution between the two carbon sublattices explains this finding that
is confirmed by the analysis of the diffusion barriers. Binding energy
of F on bi-layer systems is also computed slightly smaller than on monolayer
and electronic decoupling is observed when only one of the layer is
exposed to fluorine.

\end{abstract}

\vspace{2pc}
\maketitle

Graphene is one of the most promising materials for optical and
electronic applications \cite{Castro-Neto}. One of the direction of
research to control its electronic properties is the creation of
topological defects \cite{Nair2,Huang} and the chemical
functionalization \cite{Elias,Balog,Shih,Zhang,Kumar}. A change of the
global doping level of electrons or holes could then be obtained
together with a modification of the local electronic properties,
depending on the type, on the concentration and
on the position of the defects \cite{Robinson, Nair,Lehmann,HSantos2}.

In particular, due to the covalent character of the carbon-fluorine
bonds, the F functionalization of graphene is rather easy and large
amount of experimental
\cite{Robinson,Nair,Withers1,Cheng,Stine,Felten,Jeon,Withers2,Leefluorination}
and theoretical \cite{Liu,Sofo,Sahin,Wei} works have been reported on
the nature of the fluorination of graphene.  The CF systems, with a
coverage of the two sides of graphene layer, called fluorographane,
has demonstrated thermal and chemical stability until 600
K\cite{Nair} and a bandgap around $3.8$ eV\cite{Jeon,Nair}. Potential
applications in electronic and optoelectronic devices have been
proposed\cite{Jeon}.  Other F coverage densities modify the
bandgap as well as the electronic behavior of the
samples\cite{Cheng,Withers2,Leefluorination,Liu}. Moreover,
fluorination of selective-areas of graphene has been achieved by
removing F from graphene by Electron Beam\cite{Withers2} or by local
deposition of F by Laser irradiation with Fluoropolymers
\cite{Leefluorination}.

Other authors reported a saturated F coverage at 25\% (C$_4$F) over
one side of graphene\cite{Robinson,Felten}. These results
can also be influenced by the number of layers of graphene \cite{Felten}. A
fully fluorinated bilayer graphene has also been predicted to be more
stable than pristine bilayer graphene \cite{Sivek} because the sp$^3$
character of the C-F bonds promotes C-C bonds between layers.

The nature and consequences of the absorption of fluorine on single-
and bi-layer graphene have however not been fully investigated. The
formation of sp$^3$ bonds following the F adsorption is, as a first
approximation for $\pi$-electrons, similar to the presence of
vacancies. Both give rise to localized states at or near the Dirac
point \cite{HSantos2}. The basics of the emergence of these states lies in the
presence of two sublattices in graphene \cite{Inui}. When one $p_z$
orbital is removed in an otherwise perfect lattice, a zero-energy
state appears on the other sublattice, damped with the distance
\cite{Liang,Ducastelle}. This has also been reported for Hydrogen
absorption \cite{Duplock,Soriano}. The importance of the sublattice
symmetry in chemically modified graphene has also been reported for
nitrogen substitution \cite{Lambin}. The analysis of the stability,
of the chemical bonding and of the electronic properties of fluorinated
graphene for different coverage densities and sublattice symmetries is
then necessary.

In the present work, we investigate the binding energy (BE) and
electronic properties of fluorinated single- and bi-layer graphene by
density functional theory (DFT){\it ab-initio} calculations. Our
results demonstrate that the BE strongly depends on the F coverage
density but also on the sublattice of the carbon atoms covalently
bonded with the F atoms. Indeed, BE increases for high coverage when
carbon of different sublattices are involved in the C-F bonds and BE
decreases otherwise. This result is analyzed in terms of sublattice
symmetry of the charge distribution associated with a simple defect
or adatom. The analysis of the diffusion barriers of F on graphene
reinforces the low probability of same sublattice chemisorption for
high F coverage.  The electronic properties have also been found to
depend on the symmetry of the coverage : fluorinated graphene is an
insulator when different sublattice are occupied and metallic when
only one sublattice is considered. Furthermore, magnetic behavior is
observed only for the less stable configurations, i.e., when only one
sublattice is involved. For bilayer graphene, the BE is obtained
slightly smaller than for monolayer, in agreement with experimental
data \cite{Felten} and an electronic decoupling between the carbon
layers is observed when only one of the two layer is covered with
fluorine.

\section*{Methods}

First principles calculations have been performed using the VASP code
with spin polarization \cite{VASP}. We use the van der Waals functional
parameterized by M. Dion {\it et al.} (vdW-DF) \cite{DRSLL}, which has
been implemented in the code by J. Klimes {\it et al.}
\cite{Klimes}. The factorization proposed in 
Ref. \cite{Roman&Soler} represents a very substantial efficiency
improvement in the evaluation of the exchange-correlation potential
and energy, thus enabling first-principles van der Waals calculations
for any system accessible to usual GGAs. The results presented below
have been performed using by the functional vdW-DF, but we have
checked that other van der Waals functionals implemented in the VASP
code preserve the main features found employing vdW-DF. Spin polarized
calculations normally require a fine sampling of the Brillouin zone,
that we performed with a Monkhorst-Pack scheme and the number of point
with dependence on the size of the system.  For example, 25x25x1 and
5x5x1 k-points have been considered for C$_2$F ($3$ atoms) and
C$_{50}$F ($51$ atoms), respectively.  We have verified that the
interlayer space in graphite is in agreement with previous
calculations \cite{DRSLL,HSantos1}. The cut-off energy for plane wave basis set
is $500$ eV. The structure was relaxed by conjugate gradients
optimization until forces are smaller than 0.01 eV/{\AA}.  Periodic
boundary conditions were applied, so we use large enough supercell
parameters (33 {\AA}) in the directions perpendicular to the graphene
plane to avoid spurious interactions between adjacent layers.

In order to investigate the stability of the
fluorinated coverage, we calculate, from the total energies, the
binding energy (BE) per fluorine atom as the difference between the
fluorinated graphene (CF$_x$) and an isolated graphene plus atomic
fluorine.

The surface diffusion of F on graphene have been evaluated with the nudged
elastic band method (NEB) \cite{Mills}, as implemented in VASP. This
method allows to keep the distances between the images along a path
constant to first order.

\section*{Fluorine coverage of single layer graphene}

\begin{figure}[h!] 
  \centering
        \includegraphics[clip,width=7cm,angle=0]{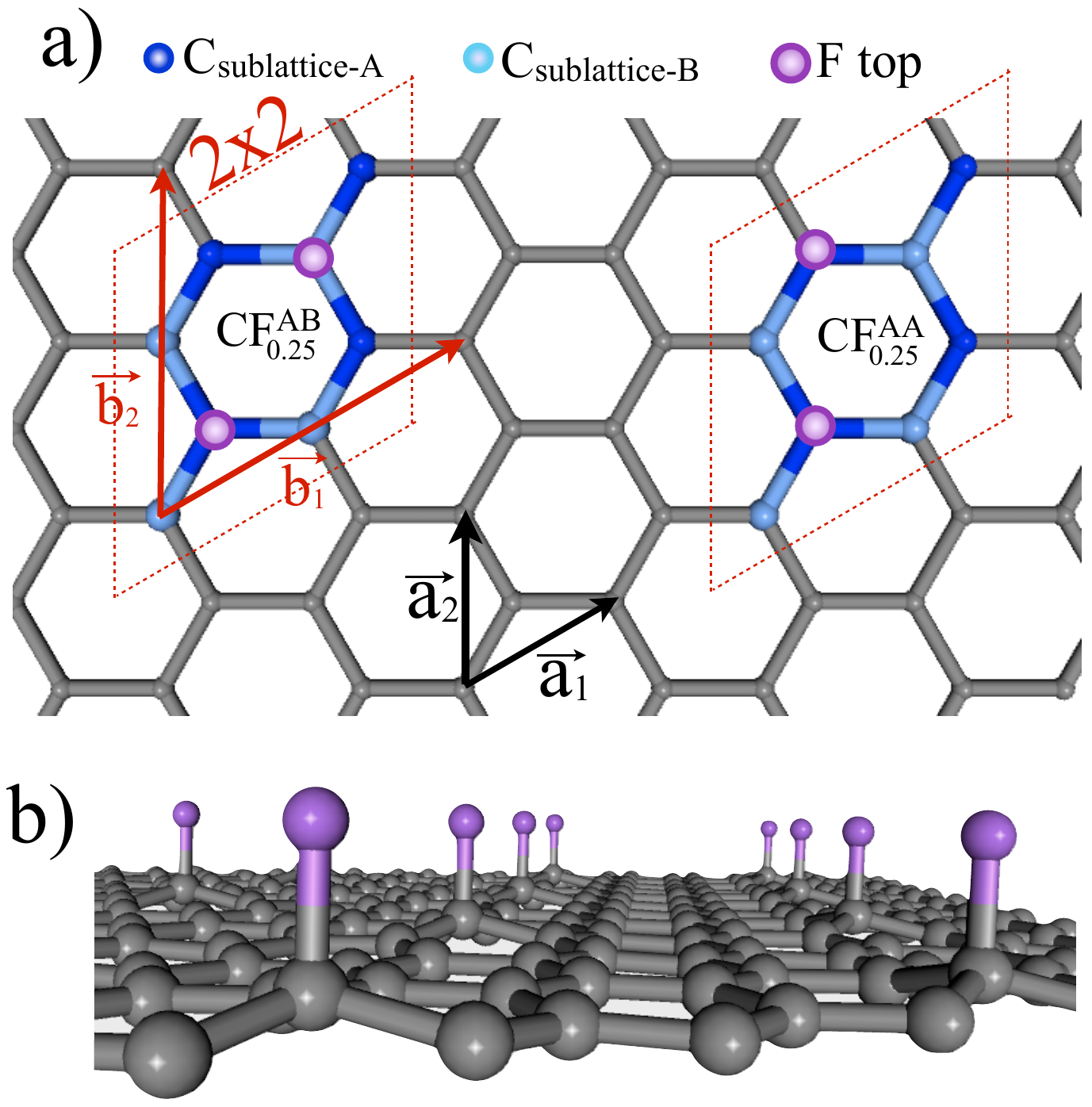}
  \caption{(Color online) a) Schematic representation of a CF$_{0.25}$
    (C$_8$F$_2$) system in a $2$x$2$ graphene supercell (red dotted
    line). Light (dark) blue balls are for the carbon atom of the
    sublattice A (B) and purple balls are for the top F atoms. Left :
    CF$_{0.25}^{AB}$. Right : CF$_{0.25}^{AA}$. b) 3D representation
    of C$_8$F$^{AA}$  in a $2$x$2$ graphene supercell.}
  \label{fig1}
 \end{figure}

In order to model the functionalization of graphene with F, we
have considered supercells of graphene denoted by $n$x$m$, where $n$ and
$m$ multiply the unit cell vectors $\vec{a}_1$ and $\vec{a}_2$,
respectively (Fig. \ref{fig1}). The number of C atoms per supercell is
$N=$$2$x$n$x$m$.  The minimum unit cell of graphene (given by $n=1$,
$m=1$) has two carbon atoms, which defined the A and B sublattices.
We have considered two important parameters for modelisation of the
fluorination of graphene : (i) the density of the coverage, (ii) the
position of the F atoms into the supercell (sublattice symmetry and
the distances between F atoms). Moreover, except otherwise stated, we
have studied systems with the F atoms lying on the same side of the
graphene in order to mimick the chemical functionalisation of graphene
deposited on a substrate.

We denote the density of coverage by $x$, in a CF$_x$ (or
C$_{n}$F$_{nx}$ system).  Lower coverages of F atoms is obtained with
large supercell, such as CF$_{0.02}$ (C$_{50}$F) generated by a
$5$x$5$ supercell. We have considered densities from $x=0.02$ to
$x=1$. The second important parameter is the position of the F
atoms. If only a single F atom is considered for a given supercell,
all the F atoms are bonded with C atoms belonging to the same
sublattice. The system is called CF$_x^{AA}$. The effect of the
sublattice symmetry can then only be investigated when several F atoms
per unit cell are considered and CF$_x^{AB}$ is used for a system with
both A and B carbon atoms chemically bonded with F atom.

The most stable system reported until now is the fluorographane (CF)
\cite{Robinson} with a F atom bonded to each C atoms, on both sides of
the graphene plane. The BE in the present van der Waals scheme
calculations is found to be BE$=2.78$ eV. In ref. \cite{Leenaerts} a
BE$= 2.86$ eV has been calculated by first principles using a
different functional (PBE). Note, two F atoms over the same side of
graphene plane in a CF system is not stable due to the repulsion
between F atoms.

As a second example, we consider C$_{2}$F$^{AA}$ with a F atom in a
$1$x$1$ supercell with all the F atoms attached to a same sublattice C
atom (say, the A sublattice). C$_{2}$F$^{AA}$ is the less stable
configuration studied in this work with a BE$=0.90$ eV. Its
counterpart, with the same concentration $x=0.5$ in the AB
configuration (C$_4$F$_2^{AB}$) in a $2$x$1$ supercell and F atoms on both sides of the
graphene layer to avoid repulsion between first neighbour F atoms has a
BE$=2.47$ eV. This means a difference of $1.57$ eV for the same
density.

C$_8$F$_2$ is obtained with a $2$x$2$ supercell with 2 F atoms (Fig.
\ref{fig1}).  When the F atoms are adsorbed on the same side but on
different sublattices (C$_8$F$_2^{AB}$) and with the larger distance
between them (equals to the third nearest neighbour), the system is the
most stable one-side' fluorinated graphene (BE = 2.27 eV), in
agreement with ref \cite{Robinson,Felten,Sofo}. The other possible
configuration, C$_8$F$_2^{AA}$, with the F atoms over the same
sublattice, is less favorable by $ 0.94$ eV. The main
structural parameter for the stability of graphene-F systems is then
the sublattice symmetry.

\begin{figure}[h!] 
  \centering
   \includegraphics[clip,width=8cm,angle=0]{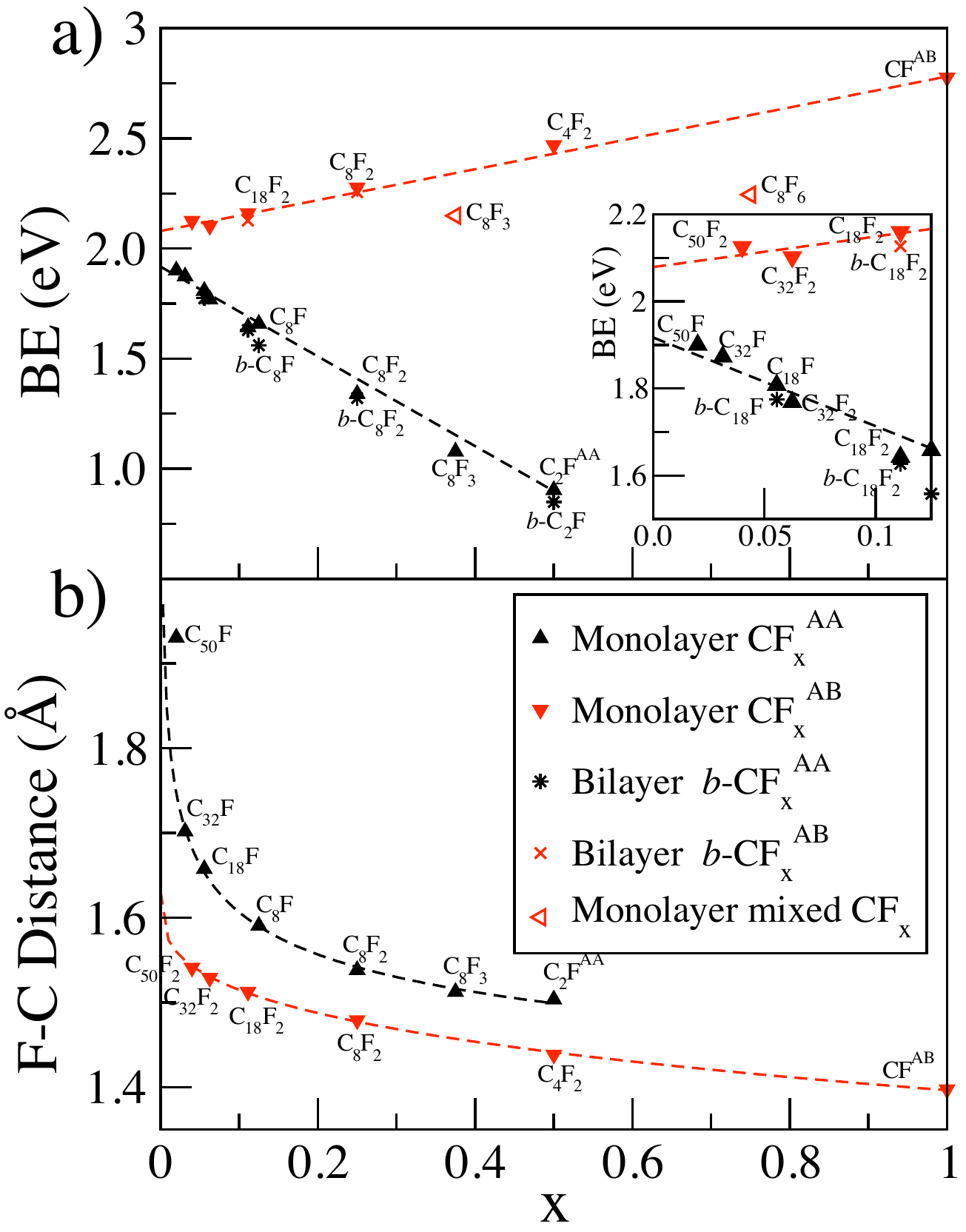}
  \caption{(Color online) Binding energy (BE) (a) and F-C distance (b) of a
    CF$_x$ as a function of the F concentration $x$ for both
    sublattice symmetry. Dashed lines are guide for the eyes.}
  \label{fig2}
 \end{figure}

The importance of the sublattice symmetry on the BE is found for all
densities of F coverage as shown in Fig. \ref{fig2}a. Note that for
the CF$_x^{AB}$ systems, the BE for the most stable position of the two
F atoms in the supercell (i.e., the third nearest neighbour) have been
reported \cite{note}. We see that the BE of A-B configuration
increases almost linearly when the F density rises. This means that,
energetically, F atoms tend to form area with high coverage. At the
opposite, for the A-A adsorption, the BE decreases almost linearly
with the F coverage. For the same coverage, the difference in the BE
between the two configurations can be considerable, from $1.57$ eV for
C$_4$F$_2$ to $0.33$ eV for C$_{32}$F$_2$. As
expected, for low density, the difference of BE vanishes and the BE is
close to $2$ eV for both configurations. Therefore, from an
experimental point of view, for low concentration, the F coverage will
not depend on the sublattice symmetry, as expected. But when the
concentration rises, chemisorption on different sublattices will be
favoured.

The importance of the sublattice symmetry is further evidence by the
special 'mixed' case of C$_8$F$_3$. When the three F atoms are
positioned on the same sublattice in a $2$x$2$ supercell, BE=1.08 eV,
following the linear behavior found previously. However, when two F
atoms are on the A sublattice and the other one in the B sublattice,
the BE rise to 2.15 eV but still below the AB curve
(Fig. \ref{fig2}a), as a consequence of the 'mixed' sublattice symmetry.

Fluorine functionalization of graphene deforms the planar structure
because of the modification of the hybridization of the C-C bounds
\cite{Sofo}. Hybridization can be investigated by
structural parameters such as the F-C distance and the C-C bond angles (for
C involved in the C-F bonds). Angles of $109^{\circ}$
($90^{\circ}$) and smaller (larger) F-C distances are expected for
perfect sp$^3$ (sp$^2$) hybridization. For CF$_x^{AB}$, distance
slightly decreases for larger coverage, from $1.54$ {\AA} for
C$_{50}$F$_2^{AB}$ to $1.40$ {\AA} for CF$^{AB}$. At the same time, a small modification 
of the angles is observed (from $103.5^{\circ}$ for C$_8$F$_2^{AB}$ 
to $101.5^{\circ}$ for C$_{50}$F$_2^{AB}$) and but no notable change of the
charge on the F atoms is observed ($0.56$ electron). Larger coverage in  CF$_x^{AB}$
is then  associated with slightly more sp$^3$ character of the C-F bonds.

\begin{figure}[htbp] 
  \centering
    \includegraphics[clip,width=10cm,angle=0]{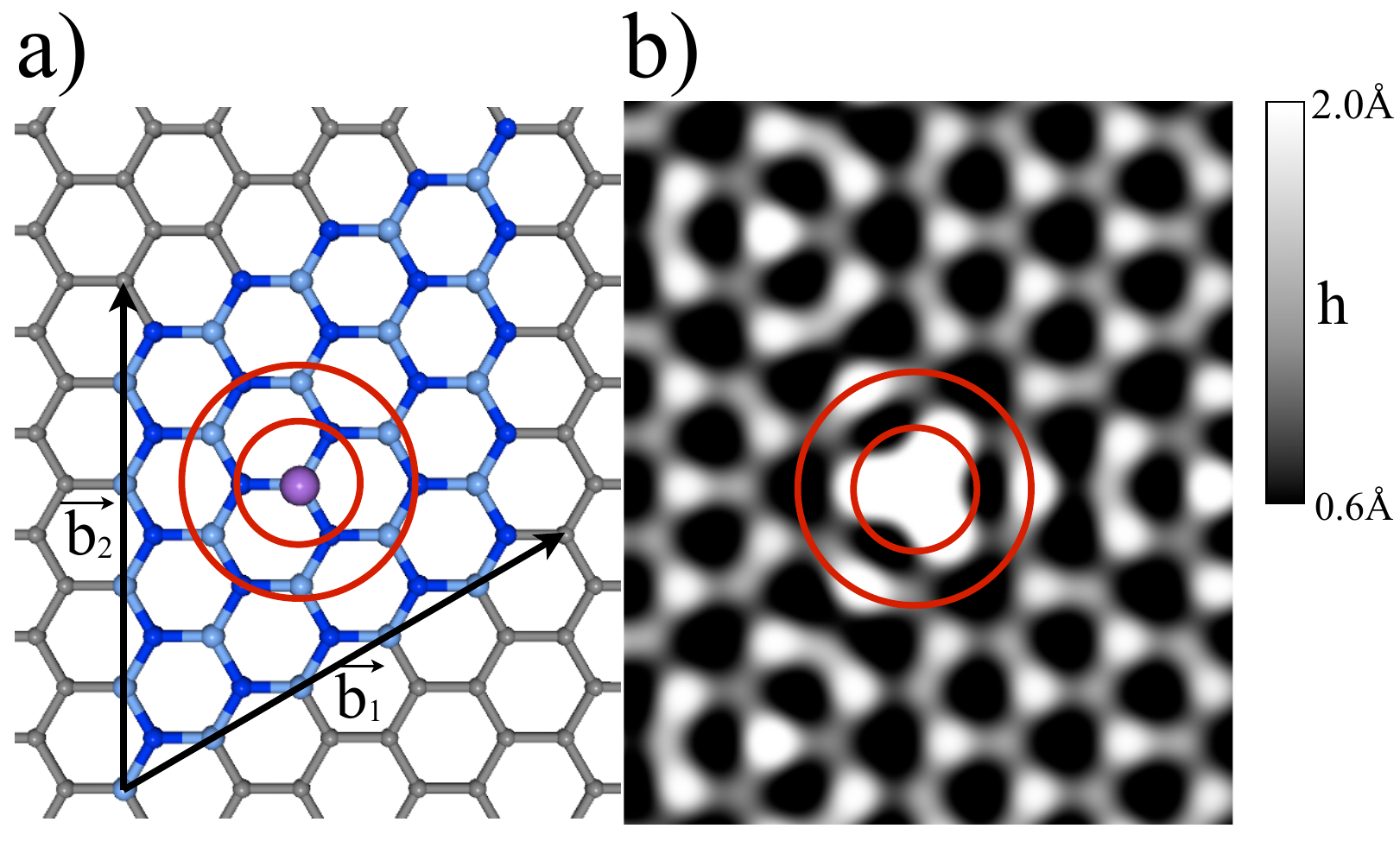}    
  \caption{(Color online) a) C$_{50}$F system. $\vec{b}_1$ and
    $\vec{b}_2$ indicate the supercell vector in the graphene
    plane. b) simulated STM image. The inner and the
    outside red circles indicate first and third neighbour from F-C
    bond, respectively.  }
  \label{fig2A}
 \end{figure}

For CF$_x^{AA}$, the C-F distance also decreases with the coverage
(Fig. \ref{fig2A}b). This is associated with a more pronounced change
of the angle from $101.4^{\circ}$ (C$_{2}$F$^{AA}$) to $96.3^{\circ}$
(C$_{50}$F$^{AA}$) and of a charge transfer varying from 0.47 electron
(C$_{2}$F$^{AA}$) to 0.62 electron (C$_{50}$F$^{AA}$). For CF$_x^{AA}$
also, larger coverage is associated with more sp$^3$ character of the
C-F bonds. However, at low coverage the C-F bonds is more ionic
and the C-C bonds are sp$^2$. The presence of two kinds of C-F
bonds (one ionic associted with sp$^2$ and one more covalent
associated with sp$^3$ have been observed experimentally by XPS
measurement \cite{Felten}.

An isolated chemical doping or defect breaks the sublattice symmetry
as demonstrated for vacancies \cite{Ducastelle}, nitrogen substitution
\cite{Lambin} or H chemisorption \cite{Duplock,Soriano}. In order to
demonstrate this feature for F chemisorption on graphene and
interprete the difference observed above for the two systems, we
present on Fig. \ref{fig2A} a simulated STM images of
C$_{50}$F$^{AA}$ ($x=0.02$) in the Tersoff-Hamman approximation for
the electronic states between $E_F$ and $E_F + 1 eV$.  Note that this
image has been computed for the 'free' side of the graphene in order to
avoid the direct imaging of the F atom. Fig.\ref{fig2A} then shows the
carbon local electronic density of states near the Fermi level for a
C$_{50}$F$^{AA}$ system. The enhanced electronic density on the B sublattice
is clearly evidenced (F atom is on the A sublattice). The formation
of a sp$^3$ bond on one sublattice then induces an increase of the electron
density (near the Fermi level) on the other sublattice, as for
vacancies. A second electronegative F atom then preferentially form
covalent sp$^3$ bonds with carbon atoms of the other (B) sublattice,
presenting an excess of electrons. This effect vanish with the
distance between two F atoms because of the screening and smaller
distance between F atoms (larger coverage) is favored. If the
same (A) sublattice is considered, the low electron density will
favor a ionic C-F bonds and sp$^2$ C-C bonds.

Following this argument, for the second F atom, the first neigbours
A-B functionalisation should be more stable. But the F-F repulsion excludes
this possibility. We have further checked the role of the F-F repulsion for the
C$_8$F$_{2}^{AB}$. When the two F atoms are positioned on the same side as
first nearest neighbour, the BE is $1.80$ eV. But when the F
atoms are chemisorbed on different side of the graphene, the BE rises
to $2.42$ eV per F atom. This large different energy (0.62 eV per
F atom) is mainly related to the repulsion between the F atom. This is
further proved if we analyze the same C$_8$F$_{2}^{AB}$ system but with two F
atoms at third nearest neighbour position. In this case,
BE=$2.27$ eV for the F atoms lying on the same side of the graphene
and BE = $2.28$ eV for the F atoms on both sides of graphene. The small
difference of energy in this last case demonstrates the preponderance
of the charge distribution on the graphene sublattice for the analysis
of the BE.

\section*{Diffusion barrier} 

Beside the total energy calculations, diffusion of F atoms on the
graphene surface is influenced by the F coverage. We use the nudged
elastic band (NEB) method to evaluate the diffusion barrier of F atoms
on graphene.  The diffusion barrier of an isolated F on graphene is found to be
$356.4$ meV for a 2x2 supercell. This very high barrier (more
than 10 times the thermal energy) is expected because of the covalent
C-F bonds. F atoms are then not mobile on a graphene surface. 

\begin{figure}[htbp] 
  \centering
        \includegraphics[clip,width=10cm,angle=0]{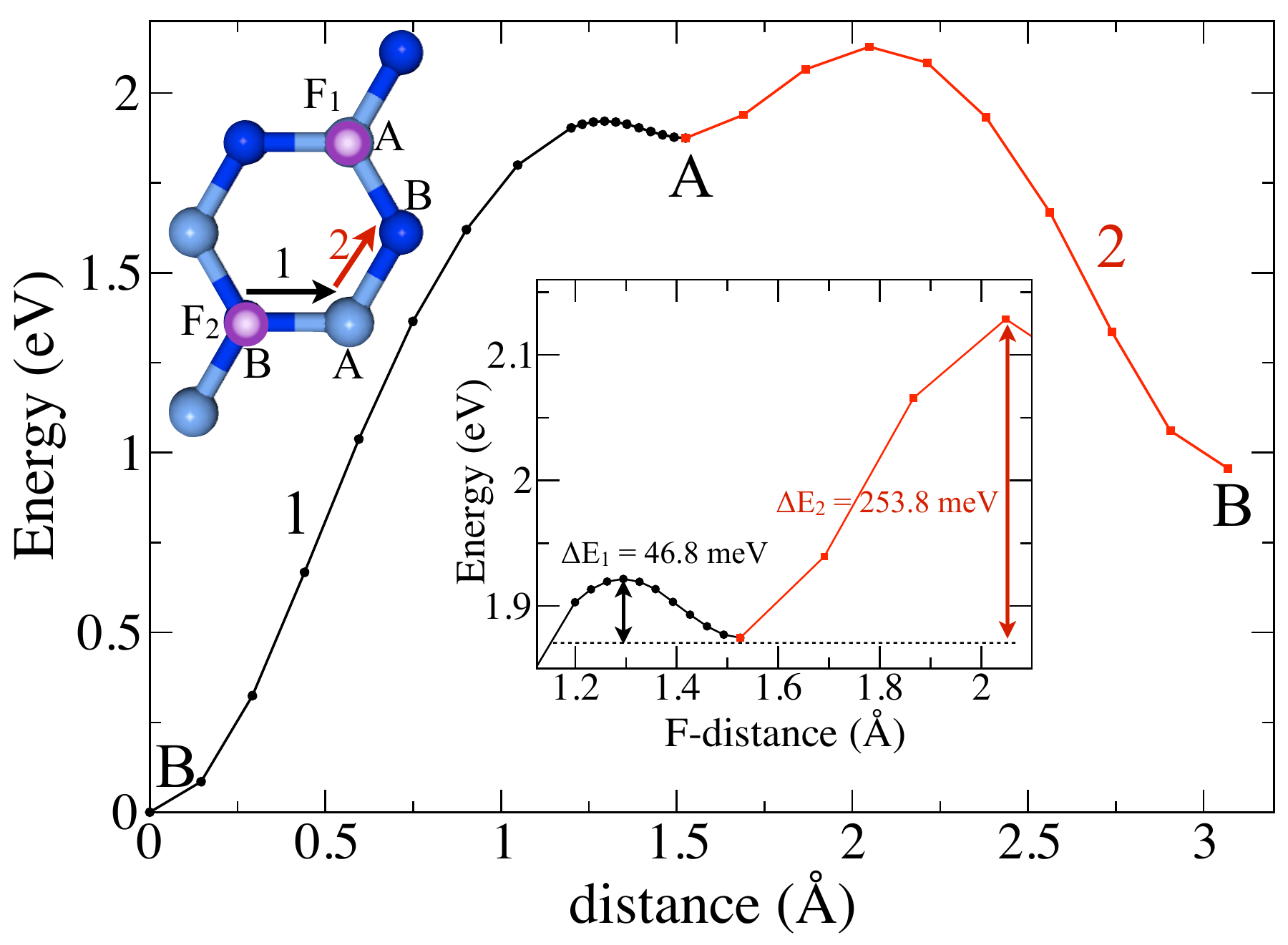}
  \caption{(Color online) Energy barrier in a C$_8$F$_{2}$ system when
    a F$_2$ atom moves along the paths $1$ (black line) and $2$ (red line).}
  \label{fig3}
 \end{figure}

The picture changes completely if a F atom is
already chemisorbed on graphene. Energy barrier in a 2x2 supercell
with 2 F atoms (C$_8$F$_{2}$) is displayed on Fig. \ref{fig3}. As
discussed before, the most stable situation is found for a F atom
bonded to C atoms of different sublattices but not first neighbours (F1
and F2 positions on fig. \ref{fig3}). To move one F atom
from the stable position to the neighbouring A site (path 1 from B to
A on Fig. \ref{fig3}), the F2 atom has to overcome a barrier that is larger than
2 eV and that correspond more or less to the difference of binding
energy between the structures. Interestingly, despite the covalent C-F
bonding, the reverse path has a barrier of only $47$ meV. This process is likely to
occurs thermally. The diffusion to the first-neighbour position (path
2) is very unlikely even if the final position is more
stable (diffusion barrier $254$ meV).

We note here that if the two F atoms are first neighbour, in spite of
the repulsion between the two F atom, this configuration is more
stable than the A-A configuration and the energy barrier along path 2
(From B to A) in more than 1 eV. However, this situation is very
unlikely to occurs because in a deposition process first neighbour
positions will be prevent by the F-F repulsion before C-F bonds can be formed.

\section*{Electronic properties}

\begin{figure}[h] 
  \centering
       \includegraphics[clip,width=15cm,angle=0]{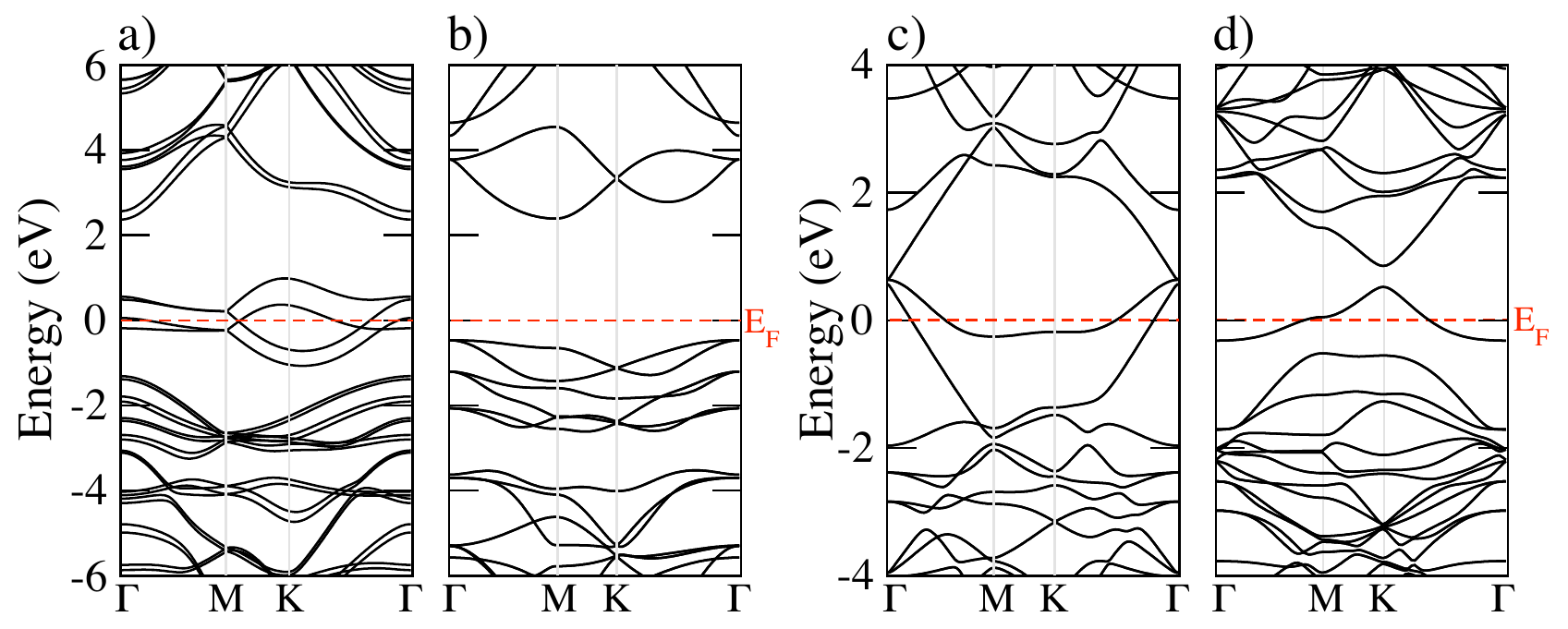}
  \caption{Band structures of a) C$_8$F$_2^{AA}$, b) C$_8$F$_2^{AB}$, c) C$_{18}$F$^{AA}$, and d) C$_{32}$F$^{AA}$. Red dashed line corresponds the Fermi level (E$_F$).}
  \label{fig4}
 \end{figure}

We now turn to the analysis of the the band structures (BS) of the C-F
systems. For single-layer graphene, as for the BE, the electronic
behaviour depends strongly on the arrangement of the F atoms into the
unit cell.  For high F concentration the Dirac cone of the pristine
graphene disappears for both F sublattices arrangements (Ref
\cite{Liu} and Fig. \ref{fig4}). For exemple, C$_8$F$_{2}^{AA}$ has almost
flat bands crosses the Fermi level (Fig. \ref{fig4}a) and the local density
of states (not shown) shows, as expected, a localization of the
associated wavefunctions on the other (B) sublattice. The system is magnetic with a
magnetic moment of $0.799 \mu _B$. For the other (more stable)
configuration, the CF$_{0.25}^{AB}$ (Fig. \ref{fig4}b), the system is
non-magnetic and semiconducting, with a gap of $2.87$ eV, in agrement
with \cite{Liu}.  For larger coverage, C$_2$F$^{AA}$ system is
metallic and has a magnetic moment of $0.745 \mu_B$, while
C$_4$F$_2^{AB}$ is non-magnetic and presents a $1.84$ eV gap.

For lower F concentration, the Dirac cone can be visualized and the
magnetisation tends to zero. For AA configurations, difference also
appears depending of the symmetry of the graphene $n$x$n$ supercell. As
for nitrogen \cite{Lambin}, if $n$ is a multiple of three
(Fig. \ref{fig4}c for the $3$x$3$ system), no gap is created. In the
other case, the Dirac cone is opened at the K point of the Bruillouin
Zone (Fig. \ref{fig4}d for the $4$x$4$ system). We also note, the
p-doping of the graphene layer (the Dirac energy is above the Fermi energy) associated with the electronegativity of the Fluorine atoms.

 \section*{Bilayer graphene} 

For bilayer graphene, denoted as {\it b}-C$_{n}$F$_{nx}$, we have
only considered the most stable Bernal (AB) stacking and F atoms
bonded only to one of the two layers, in order to mimick the chemical
functionalisation of graphene lying on a substrate. Because of the
Bernal stacking, the symmetry of the sublattice is broken and the two
sublattices correspond to inequivalent absortion sites. The different
in total energy calculation between a F chemisorbed on the A (carbon
atom on the top of a carbon atom of the other layer) and B (carbon
atom of the top of the center of an hexagon) sublattices decays with
the size of the system, and is in favor of the B sublattice. This difference
is however much smaller than the one associated with the lattice
assymmetry ($9.4$, $2.5$ and $0.4$ meV per F atom for {\it b}-C$_2$F,
{\it b}-C$_8$F and {\it b}-C$_{18}$F, respectively). Note the
difference found in Ref. \cite{Sivek} is $0.3$ meV but with a PBE
functional and a F coverage for the two sides of bilayer graphene.

The BE for {\it b}-CF$_x$, with F atoms chemisorbed on one of
the layers only, follows the trends observed for single-layer graphene
(Fig. \ref{fig2}).  {\it b}-CF$_x$ however exhibits systematically
smaller BE than monolayer CF$_x$. The difference is however very small
( $0.033$ eV for C$_{18}$F$^{AA}$) and is barely visible on the
figure. A fluorination of a bilayer graphene energetically less
favorable than for monolayer is on agreement with the recent
experimental data of Felten {\it et al.}  \cite{Felten} even if the
present order of magnetitude is too small to explain the
observations. Note also that the difference between {\it
  b}-C$_8$F$_{2}^{BB}$ and {\it b}-C$_8$F$_2^{AB}$ is of 0.94 eV in
favor of the AB case, similar to the diffrence for single layer graphene
(Fig. \ref{fig2}).

Charge calculations indicate a weak charge
transfer from the pristine layer to the functionalized layer of 0.0043 e
per C atom for {\it b}-C$_2$F$^{AA}$ system, 0.0029 e per C atom for {\it
  b}-C$_8$F$^{AA}$. 

Fluorination also influences the interlayer interaction energy (ILE)
per carbon atom. For example, ILE are found to be $0.016$ eV and
$0.021$ eV for {\it b}-C$_2$F{$^{AA}$} and {\it b}-C$_2$F{$^{BB}$},
respectively, compared with $0.050$ eV for pristine bi-layer
\cite{HSantos1}. For more stable C$_8$F$_2^{AA}$, C$_8$F$_2^{BB}$ and
C$_8$F$_2^{AB}$ ILE is even smaller ($0.015$, $0.001$ and $0.002$ eV,
respectively). A decoupling between the graphene layers then occurs
due to the fluorination and is correlated with the sp$^3$
hybridization of one of the layers.  A decoupling in bilayer graphene
has also been observed \cite{Felten2} and simulated for oxygen
chemisorption \cite{Nourbakshs}. At the opposite, a fluorination of
both side of bilayer graphene lead to a strong coupling due to the
formation of sp$^3$ bonds between C atoms of different layers
\cite{Sivek}.

\begin{figure}[t] 
  \centering
   \includegraphics[clip,width=14.cm,angle=0]{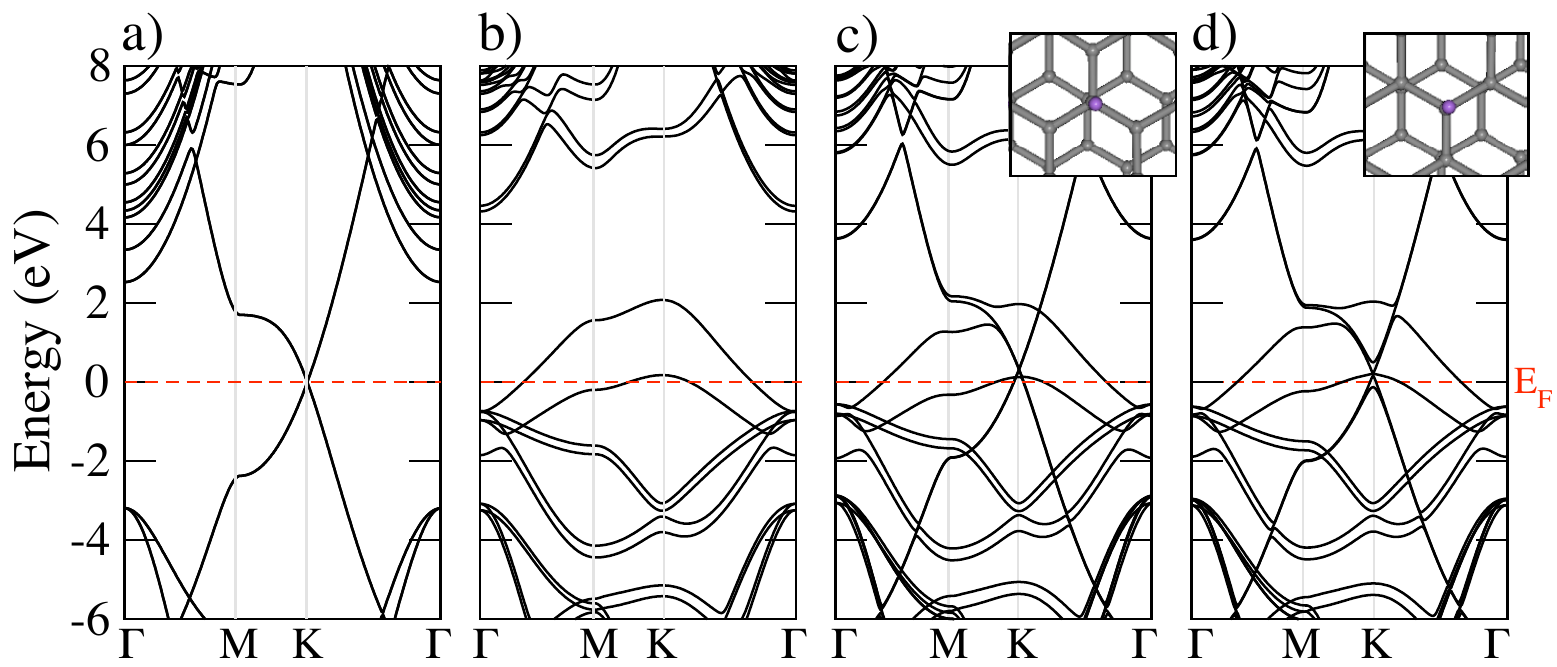}        
  \caption{ (Color online) Band structure of a) graphene $1$x$1$, b)
    C$_2$F$^{AA}$, c) {\it b}-C$_{2}$F$^{AA}$, and d) {\it
      b}-C$_{2}$F$^{BB}$. Red dashed line corresponds the Fermi level (E$_F$).
     }
  \label{fig5}
 \end{figure}

The electronic band structures (BS) of the systems also demonstrate the
decoupling between the two layers when one of them is fluorinated.
Fig. \ref{fig5} c and d show the BS of a {\it b}-C$_2$F when the F
atom is bonded to the A and B sublattices, respectively.  For
comparison, the BS of pristine graphene single layer (Fig. \ref{fig5}a) and
single layer C$_2$F$^{AA}$ (Fig. \ref{fig5}b) are also presented . The
BS of the {\it b}-C$_2$F$^{AA}$ system is almost the sum of two
isolated systems. The fluorination of one of the layer of bi-layer
graphene then results in an electronic decoupling of the two layers
and Dirac fermion behavior is recovered on the pristine layer. The
Dirac cone is slighly shifted up due to the p-doping mentionned
earlier. If we closely look at the Dirac point, the electronic
decoupling is better for C$_2$F$^{AA}$, as expected from the ILE
results.  Similar results have been obtained for smaller coverage,
with a smaller decoupling as the coverage decreases. (see
Suppl. Information). Nevertheless, for the energetically favourable
{\it b}-C$_8$F$_{2}^{AB}$ the decoupling is still
strong.

 \section*{Summary} 
 
In conclusion, the nature of the F functionalization of graphene and
bilayer graphene have been addressed. The study by means of van der
Waals first-principles calculations demonstrate the sublattice
dependence in chemisorption in graphene. The most stable F coverages
are obtained when the two sublattices are equally occupied by F
atom. Combined with the short range F-F repulsion, the most stable
system is C$_4$F, as observed experimentally
\cite{Robinson,Felten,Jeon}. Energy barriers calculations reinforced
this conclusion. Bilayer flurorination is slightly less favourable
energetically but present an interesting electronic decoupling between
graphene layer. This decoupling could be observabled experimentally
by STM, Raman \cite{ChaconTorres,Felten2}, transport measurements or
ARPES. Chemisorption of top layer bilayer graphene can be a way to (locally)
create an electronically single-layer graphene sandwiched between an
insulating F functionalized graphene layer and the substrate.

\section*{Acknowledgments}
The authors thank Thomas Chanier for helpfull advices on NEB
calculations and Alexandre Felten usefull discussions on experimental data. This work was
supported by Spanish Ministry of Economy and Competitiveness (MINECO)
under Grants No. FIS2010-21282-C02-02 and used resources of the
"Plateforme Technologique de Calcul Intensif (PTCI)"
(http://www.ptci.unamur.be) located at the University of Namur,
Belgium, which is supported by the F.R.S.-FNRS under the convention
No. 2.4520.11. The PTCI is member of the "Consortium des Équipements
de Calcul Intensif (CÉCI)"
(http://www.ceci-hpc.be). H. S. acknowledges the F.R.S.-FNRS for a
mobility grant and the hospitality of the Department of Physics of the
University of Namur.

\section*{References}
\bibliography{bib_Fluorine}

\providecommand{\newblock}{}
\begin{thebibliography}{10}
\expandafter\ifx\csname url\endcsname\relax
  \def\url#1{{\tt #1}}\fi
\expandafter\ifx\csname urlprefix\endcsname\relax\def\urlprefix{URL }\fi
\providecommand{\eprint}[2][]{\url{#2}}

\bibitem{Castro-Neto}
Castro~Neto A~H, Guinea F, Peres N~M~R, Novoselov K~S and Geim A~K 2009 {\em
  Rev. Mod. Phys.\/} {\bf 81}(1) 109--162
  \urlprefix\url{http://link.aps.org/doi/10.1103/RevModPhys.81.109}

\bibitem{Nair2}
Nair R~R, Tsai I~L, Sepioni M, Lehtinen O, Keinonen J, Krasheninnikov A~V,
  Castro~Neto A~H, Katsnelson M~I, Geim A~K and Grigorieva I~V 2013 {\em Nat
  Commun\/} {\bf 4} \urlprefix\url{http://dx.doi.org/10.1038/ncomms3010}

\bibitem{Huang}
Huang P~Y, Ruiz-Vargas C~S, van~der Zande A~M, Whitney W~S, Levendorf M~P,
  Kevek J~W, Garg S, Alden J~S, Hustedt C~J, Zhu Y, Park J, McEuen P~L and
  Muller D~A 2011 {\em Nature\/} {\bf 469} 389--392
  \urlprefix\url{http://dx.doi.org/10.1038/nature09718}

\bibitem{Elias}
Elias D~C, Nair R~R, Mohiuddin T~M~G, Morozov S~V, Blake P, Halsall M~P,
  Ferrari A~C, Boukhvalov D~W, Katsnelson M~I, Geim A~K and Novoselov K~S 2009
  {\em Science\/} {\bf 323} 610--613 (\textit{Preprint}
  \eprint{http://www.sciencemag.org/content/323/5914/610.full.pdf})
  \urlprefix\url{http://www.sciencemag.org/content/323/5914/610.abstract}

\bibitem{Balog}
Balog R, Andersen M, Jørgensen B, Sljivancanin Z, Hammer B, Baraldi A,
  Larciprete R, Hofmann P, Hornekær L and Lizzit S 2013 {\em ACS Nano\/} {\bf
  7} 3823--3832 (\textit{Preprint}
  \eprint{http://pubs.acs.org/doi/pdf/10.1021/nn400780x})
  \urlprefix\url{http://pubs.acs.org/doi/abs/10.1021/nn400780x}

\bibitem{Shih}
Shih C~J, Wang Q~H, Jin Z, Paulus G~L~C, Blankschtein D, Jarillo-Herrero P and
  Strano M~S 2013 {\em Nano Letters\/} {\bf 13} 809--817 (\textit{Preprint}
  \eprint{http://pubs.acs.org/doi/pdf/10.1021/nl304632e})
  \urlprefix\url{http://pubs.acs.org/doi/abs/10.1021/nl304632e}

\bibitem{Zhang}
Zhang H, Bekyarova E, Huang J~W, Zhao Z, Bao W, Wang F, Haddon R~C and Lau C~N
  2011 {\em Nano Letters\/} {\bf 11} 4047--4051 (\textit{Preprint}
  \eprint{http://pubs.acs.org/doi/pdf/10.1021/nl200803q})
  \urlprefix\url{http://pubs.acs.org/doi/abs/10.1021/nl200803q}

\bibitem{Kumar}
Kumar P~V, Bernardi M and Grossman J~C 2013 {\em ACS Nano\/} {\bf 7} 1638--1645
  (\textit{Preprint} \eprint{http://pubs.acs.org/doi/pdf/10.1021/nn305507p})
  \urlprefix\url{http://pubs.acs.org/doi/abs/10.1021/nn305507p}

\bibitem{Robinson}
Robinson J~T, Burgess J~S, Junkermeier C~E, Badescu S~C, Reinecke T~L, Perkins
  F~K, Zalalutdniov M~K, Baldwin J~W, Culbertson J~C, Sheehan P~E and Snow E~S
  2010 {\em Nano Letters\/} {\bf 10} 3001--3005 (\textit{Preprint}
  \eprint{http://pubs.acs.org/doi/pdf/10.1021/nl101437p})
  \urlprefix\url{http://pubs.acs.org/doi/abs/10.1021/nl101437p}

\bibitem{Nair}
Nair R~R, Ren W, Jalil R, Riaz I, Kravets V~G, Britnell L, Blake P, Schedin F,
  Mayorov A~S, Yuan S, Katsnelson M~I, Cheng H~M, Strupinski W, Bulusheva L~G,
  Okotrub A~V, Grigorieva I~V, Grigorenko A~N, Novoselov K~S and Geim A~K 2010
  {\em Small\/} {\bf 6} 2773--2773 ISSN 1613-6829
  \urlprefix\url{http://dx.doi.org/10.1002/smll.201090086}

\bibitem{Lehmann}
Lehmann T, Ryndyk D~A and Cuniberti G 2013 {\em Phys. Rev. B\/} {\bf 88}(12)
  125420 \urlprefix\url{http://link.aps.org/doi/10.1103/PhysRevB.88.125420}

\bibitem{HSantos2}
Santos H, Soriano D and Palacios J~J 2014 {\em Phys. Rev. B\/} {\bf 89}(19)
  195416 \urlprefix\url{http://link.aps.org/doi/10.1103/PhysRevB.89.195416}

\bibitem{Withers1}
Withers F, Dubois M and Savchenko A~K 2010 {\em Phys. Rev. B\/} {\bf 82}(7)
  073403 \urlprefix\url{http://link.aps.org/doi/10.1103/PhysRevB.82.073403}

\bibitem{Cheng}
Cheng S~H, Zou K, Okino F, Gutierrez H~R, Gupta A, Shen N, Eklund P~C, Sofo J~O
  and Zhu J 2010 {\em Phys. Rev. B\/} {\bf 81}(20) 205435
  \urlprefix\url{http://link.aps.org/doi/10.1103/PhysRevB.81.205435}

\bibitem{Stine}
Stine R, Lee W~K, Whitener K~E, Robinson J~T and Sheehan P~E 2013 {\em Nano
  Letters\/} {\bf 13} 4311--4316 (\textit{Preprint}
  \eprint{http://pubs.acs.org/doi/pdf/10.1021/nl4021039})
  \urlprefix\url{http://pubs.acs.org/doi/abs/10.1021/nl4021039}

\bibitem{Felten}
Felten A, Eckmann A, Pireaux J~J, Krupke R and Casiraghi C 2013 {\em
  Nanotechnology\/} {\bf 24} 355705
  \urlprefix\url{http://stacks.iop.org/0957-4484/24/i=35/a=355705}

\bibitem{Jeon}
Jeon K~J, Lee Z, Pollak E, Moreschini L, Bostwick A, Park C~M, Mendelsberg R,
  Radmilovic V, Kostecki R, Richardson T~J and Rotenberg E 2011 {\em ACS
  Nano\/} {\bf 5} 1042--1046 (\textit{Preprint}
  \eprint{http://pubs.acs.org/doi/pdf/10.1021/nn1025274})
  \urlprefix\url{http://pubs.acs.org/doi/abs/10.1021/nn1025274}

\bibitem{Withers2}
Withers F, Bointon T~H, Dubois M, Russo S and Craciun M~F 2011 {\em Nano
  Letters\/} {\bf 11} 3912--3916 (\textit{Preprint}
  \eprint{http://pubs.acs.org/doi/pdf/10.1021/nl2020697})
  \urlprefix\url{http://pubs.acs.org/doi/abs/10.1021/nl2020697}

\bibitem{Leefluorination}
Lee W~H, Suk J~W, Chou H, Lee J, Hao Y, Wu Y, Piner R, Akinwande D, Kim K~S and
  Ruoff R~S 2012 {\em Nano Letters\/} {\bf 12} 2374--2378 (\textit{Preprint}
  \eprint{http://pubs.acs.org/doi/pdf/10.1021/nl300346j})
  \urlprefix\url{http://pubs.acs.org/doi/abs/10.1021/nl300346j}

\bibitem{Liu}
Liu H~Y, Hou Z~F, Hu C~H, Yang Y and Zhu Z~Z 2012 {\em The Journal of Physical
  Chemistry C\/} {\bf 116} 18193--18201 (\textit{Preprint}
  \eprint{http://pubs.acs.org/doi/pdf/10.1021/jp303279r})
  \urlprefix\url{http://pubs.acs.org/doi/abs/10.1021/jp303279r}

\bibitem{Sofo}
Sofo J~O, Suarez A~M, Usaj G, Cornaglia P~S, Hern\'andez-Nieves A~D and
  Balseiro C~A 2011 {\em Phys. Rev. B\/} {\bf 83}(8) 081411
  \urlprefix\url{http://link.aps.org/doi/10.1103/PhysRevB.83.081411}

\bibitem{Sahin}
\ifmmode~\mbox{\c{S}}\else \c{S}\fi{}ahin H, Topsakal M and Ciraci S 2011 {\em
  Phys. Rev. B\/} {\bf 83}(11) 115432
  \urlprefix\url{http://link.aps.org/doi/10.1103/PhysRevB.83.115432}

\bibitem{Wei}
Wei W and Jacob T 2013 {\em Phys. Rev. B\/} {\bf 87}(11) 115431
  \urlprefix\url{http://link.aps.org/doi/10.1103/PhysRevB.87.115431}

\bibitem{Sivek}
Sivek J, Leenaerts O, Partoens B and Peeters F~M 2012 {\em The Journal of
  Physical Chemistry C\/} {\bf 116} 19240--19245 (\textit{Preprint}
  \eprint{http://pubs.acs.org/doi/pdf/10.1021/jp3027012})
  \urlprefix\url{http://pubs.acs.org/doi/abs/10.1021/jp3027012}

\bibitem{Inui}
Inui M, Trugman S~A and Abrahams E 1994 {\em Phys. Rev. B\/} {\bf 49}
  3190--3196

\bibitem{Liang}
Liang S~Z and Sofo J~O 2012 {\em Phys. Rev. Lett.\/} {\bf 109}(25) 256601
  \urlprefix\url{http://link.aps.org/doi/10.1103/PhysRevLett.109.256601}

\bibitem{Ducastelle}
Ducastelle F 2013 {\em Physical Review B\/} {\bf 88} 075413 (\textit{Preprint}
  \eprint{http://journals.aps.org/prb/abstract/10.1103/PhysRevB.88.075413})
  \urlprefix\url{http://journals.aps.org/prb/abstract/10.1103/PhysRevB.88.0754%
13}

\bibitem{Duplock}
Duplock E~J, Scheffler M and Lindan P~J~D 2004 {\em Phys. Rev. Lett.\/} {\bf
  92}(22) 225502
  \urlprefix\url{http://link.aps.org/doi/10.1103/PhysRevLett.92.225502}

\bibitem{Soriano}
Soriano D, Leconte N, Ordej\'on P, Charlier J~C, Palacios J~J and Roche S 2011
  {\em Phys. Rev. Lett.\/} {\bf 107}(1) 016602
  \urlprefix\url{http://link.aps.org/doi/10.1103/PhysRevLett.107.016602}

\bibitem{Lambin}
Lambin P, Amara H, Ducastelle F and Henrard L 2012 {\em Phys. Rev. B\/} {\bf
  86}(4) 045448
  \urlprefix\url{http://link.aps.org/doi/10.1103/PhysRevB.86.045448}

\bibitem{VASP}
Kresse G and Furthm\"uller J 1996 {\em Phys. Rev. B\/} {\bf 54}(16)
  11169--11186
  \urlprefix\url{http://link.aps.org/doi/10.1103/PhysRevB.54.11169}

\bibitem{DRSLL}
Dion M, Rydberg H, Schr\"oder E, Langreth D~C and Lundqvist B~I 2004 {\em Phys.
  Rev. Lett.\/} {\bf 92}(24) 246401
  \urlprefix\url{http://link.aps.org/doi/10.1103/PhysRevLett.92.246401}

\bibitem{Klimes}
Klime\ifmmode~\check{s}\else \v{s}\fi{} J~c~v, Bowler D~R and Michaelides A
  2011 {\em Phys. Rev. B\/} {\bf 83}(19) 195131
  \urlprefix\url{http://link.aps.org/doi/10.1103/PhysRevB.83.195131}

\bibitem{Roman&Soler}
Rom\'an-P\'erez G and Soler J~M 2009 {\em Phys. Rev. Lett.\/} {\bf 103}(9)
  096102 \urlprefix\url{http://link.aps.org/doi/10.1103/PhysRevLett.103.096102}

\bibitem{HSantos1}
Santos H, Ayuela A, Chico L and Artacho E 2012 {\em Phys. Rev. B\/} {\bf
  85}(24) 245430
  \urlprefix\url{http://link.aps.org/doi/10.1103/PhysRevB.85.245430}

\bibitem{Mills}
Mills G, Jónsson H and Schenter G~K 1995 {\em Surface Science\/} {\bf 324} 305
  -- 337 ISSN 0039-6028
  \urlprefix\url{http://www.sciencedirect.com/science/article/pii/003960289400%
7314}

\bibitem{Leenaerts}
Leenaerts O, Peelaers H, Hern\'andez-Nieves A~D, Partoens B and Peeters F~M
  2010 {\em Phys. Rev. B\/} {\bf 82}(19) 195436
  \urlprefix\url{http://link.aps.org/doi/10.1103/PhysRevB.82.195436}

\bibitem{note}
For C$_4$F$_2^{AB}$, C$_8$F$_6^{AB}$ and CF$^{AB}$, F atoms are on both side of
  the graphene. No configuration with F atoms on the same side of the layer do
  exit without first neighbour position.

\bibitem{Felten2}
Felten A, Flavel B~S, Britnell L, Eckmann A, Louette P, Pireaux J~J, Hirtz M,
  Krupke R and Casiraghi C 2013 {\em Small\/} {\bf 9} 631--639 ISSN 1613-6829
  \urlprefix\url{http://dx.doi.org/10.1002/smll.201202214}

\bibitem{Nourbakshs}
Nourbakhsh A, Cantoro M, Klekachev A~V, Pourtois G, Vosch T, Hofkens J, van~der
  Veen M~H, Heyns M~M, De~Gendt S and Sels B~F 2011 {\em The Journal of
  Physical Chemistry C\/} {\bf 115} 16619--16624 (\textit{Preprint}
  \eprint{http://pubs.acs.org/doi/pdf/10.1021/jp203010z})
  \urlprefix\url{http://pubs.acs.org/doi/abs/10.1021/jp203010z}

\bibitem{ChaconTorres}
Chac\'on-Torres J~C, Wirtz L and Pichler T 2013 {\em ACS Nano\/} {\bf 7}
  9249--9259 (\textit{Preprint}
  \eprint{http://pubs.acs.org/doi/pdf/10.1021/nn403885k})
  \urlprefix\url{http://pubs.acs.org/doi/abs/10.1021/nn403885k}

\end{thebibliography}


\newpage
\title{Supporting information: Fluorine absorption on single and bilayer graphene: Role of sublattice and layer decoupling}

\author{Hern\'an Santos$^{1,2}$ and Luc Henrard$^2$}
\address{$^1$ Departamento de F\'{\i}sica Fundamental, Universidad Nacional de Educaci\'on a Distancia, Apartado 60141, E-28040 Madrid, Spain}
\address{$^2$ Research Center in Physics of Matter and Radiation (PMR), University of Namur, Rue de Bruxelles 61, B-5000 Namur, Belgium}
\eads{\mailto{hernan.santos@fisfun.uned.es}, \mailto{ luc.henrard@unamur.be}}

\pacs{61.48.Gh, 71.15.Mb, 73.22.Pr}
\vspace{2pc}
\noindent{\it Keywords}: Graphene, Bilayer Graphene, Fluorine Functionalization, Simulation, Electronic Properties, Binding Energy
\maketitle

\begin{figure}[h!] 
  \centering
        \includegraphics[clip,width=12cm,angle=0]{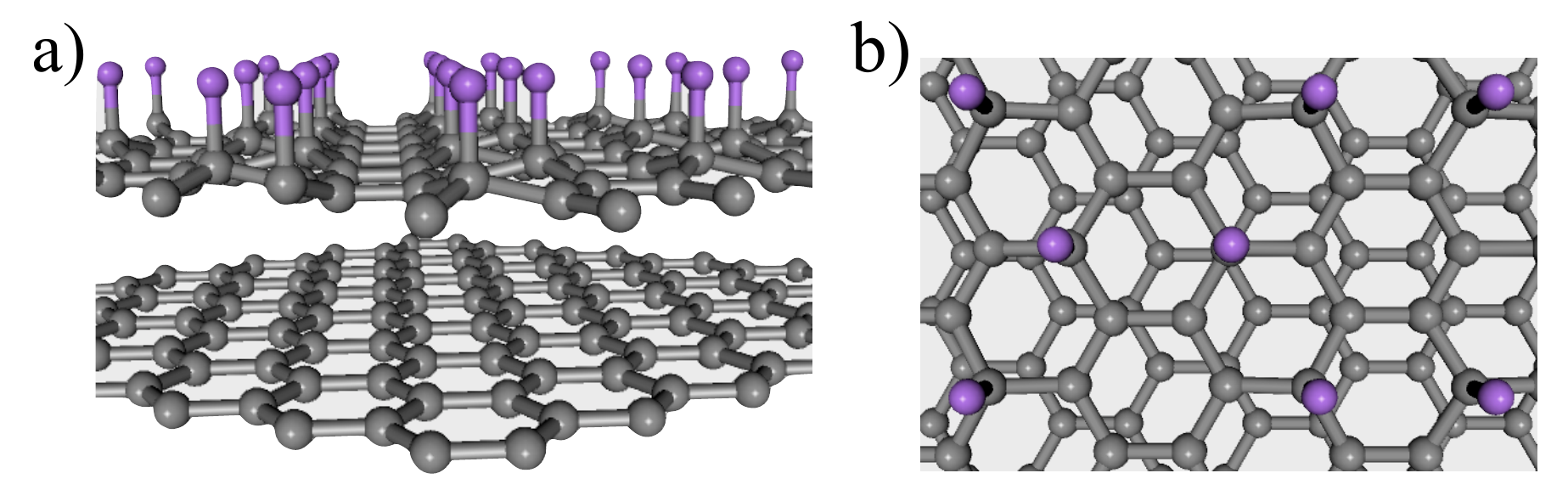}
  \caption{(Color online) 3D view of a bilayer graphene which is coveraged on the one side of one of the layers by F. The figures correspond to {\it b}-C$_{8}$F$_2^{AB}$ in a a) lateral and b) planar view.}
  \label{fig1_su}
 \end{figure}

\begin{figure}[h!] 
  \centering
        \includegraphics[clip,width=15cm,angle=0]{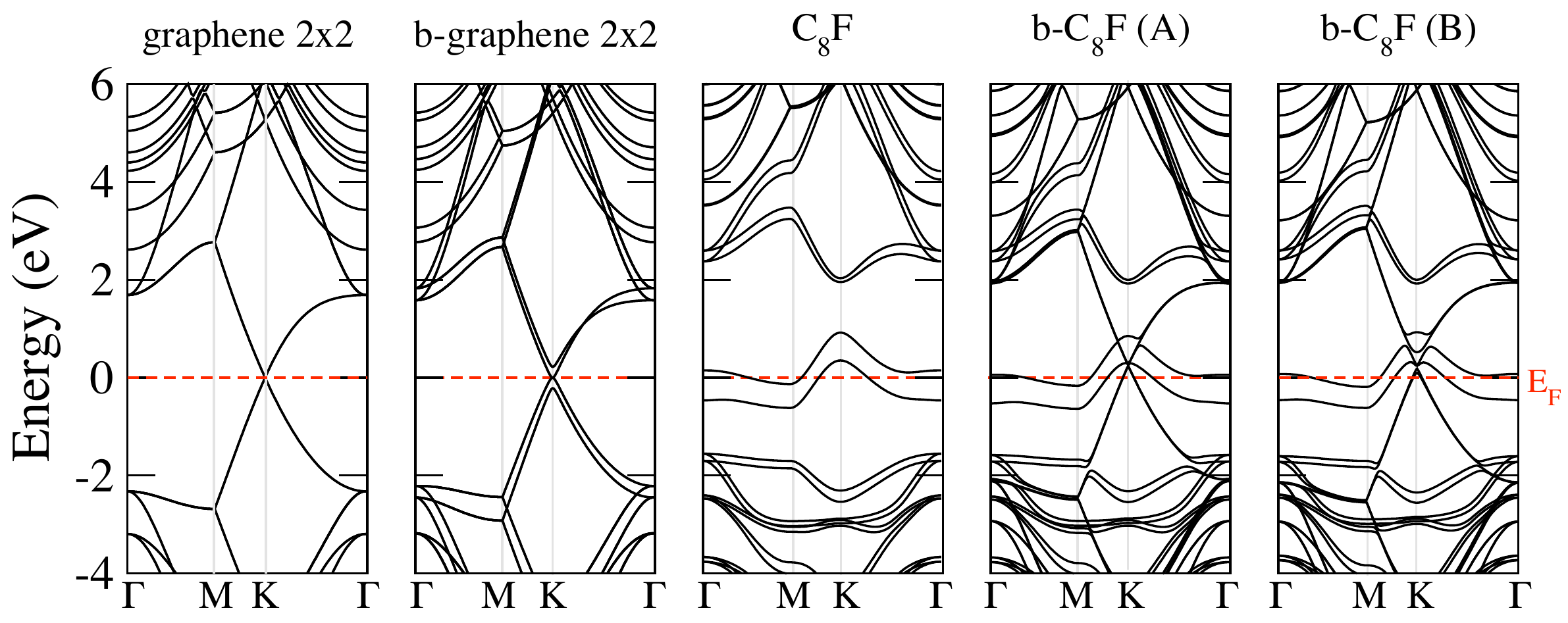}
  \caption{(Color online) Band structure of a) graphene 2x2, b) bilayer graphene 2x2, c) C$_8$F$^{AA}$, d)
  {\it b}-C$_{8}$F$^{AA}$, and e) {\it b}-C$_{8}$F$^{BB}$. Red dashed line indicates the Fermi level (E$_F$).}
  \label{fig2_su}
 \end{figure}

\begin{figure}[h!] 
  \centering
        \includegraphics[clip,width=9cm,angle=0]{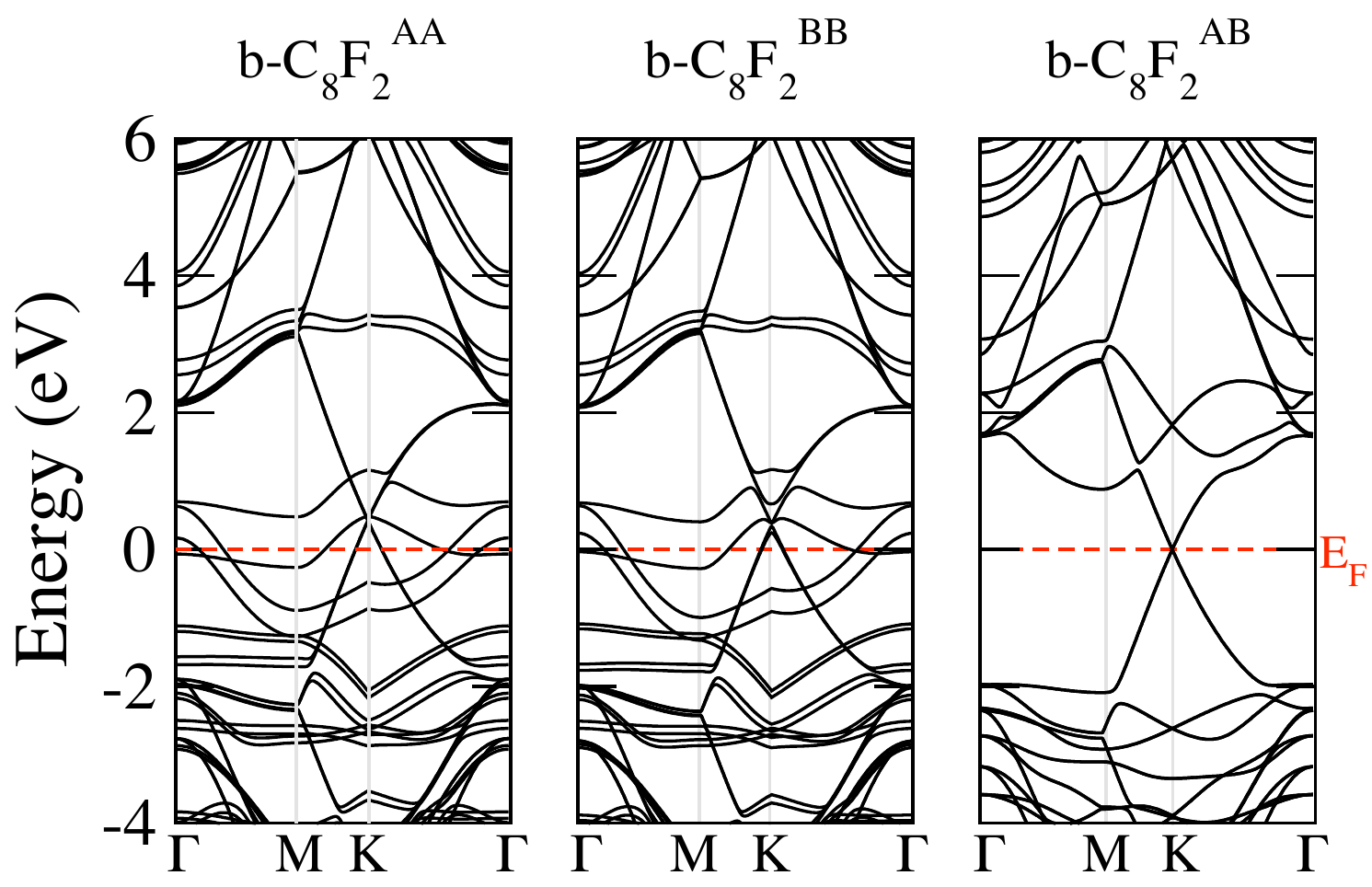}
  \caption{(Color online) Band structure of a) {\it b}-C$_8$F$_2^{AA}$, d)
  {\it b}-C$_8$F$_2^{BB}$, and e) {\it b}-C$_8$F$_2^{AB}$. Red dashed line indicates the Fermi level (E$_F$).}
  \label{fig3_su}
 \end{figure}

\begin{figure}[h!] 
  \centering
        \includegraphics[clip,width=15cm,angle=0]{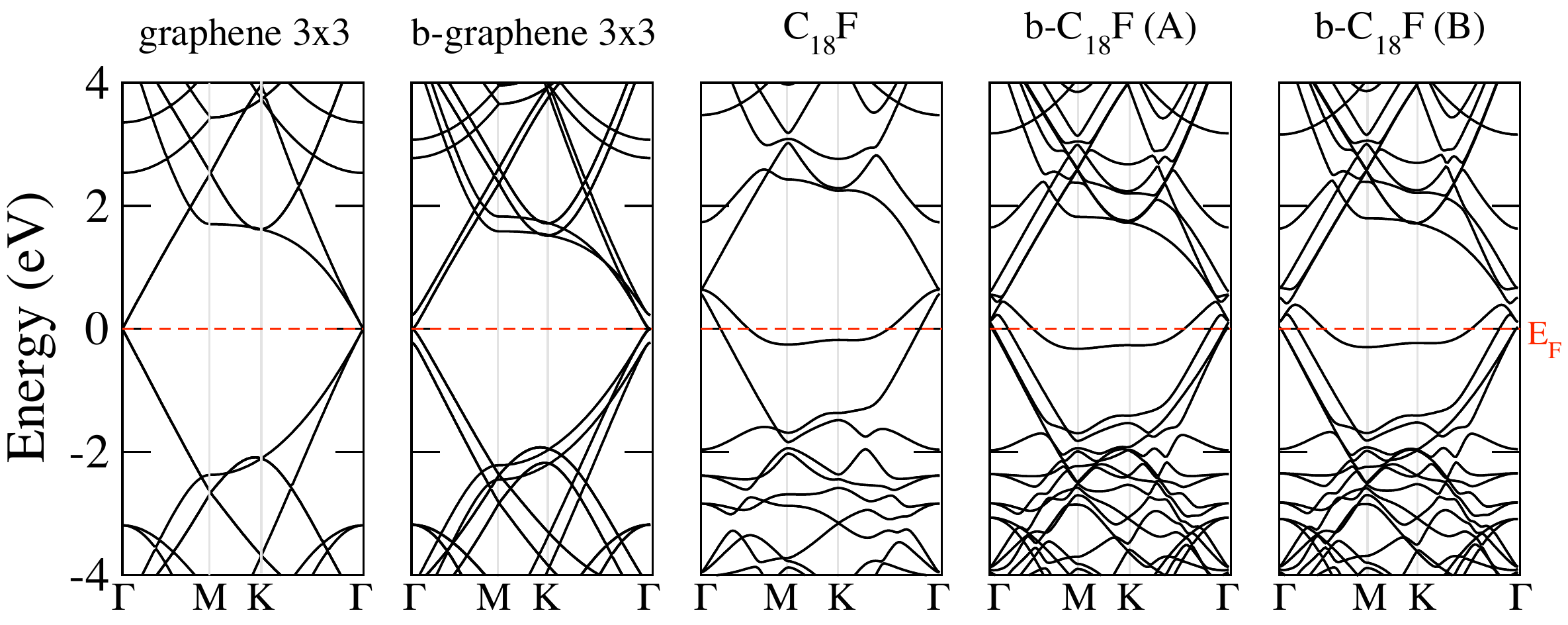}
  \caption{(Color online) Band structure of a) graphene 3x3, b) bilayer graphene 3x3, c) C$_{18}$F$^{AA}$, d)
  {\it b}-C$_{18}$F$^{AA}$, and e) {\it b}-C$_{18}$F$^{BB}$. Red dashed line indicates the Fermi level (E$_F$).  }
  \label{fig4_su}
 \end{figure}

\begin{figure}[h!] 
  \centering
        \includegraphics[clip,width=15cm,angle=0]{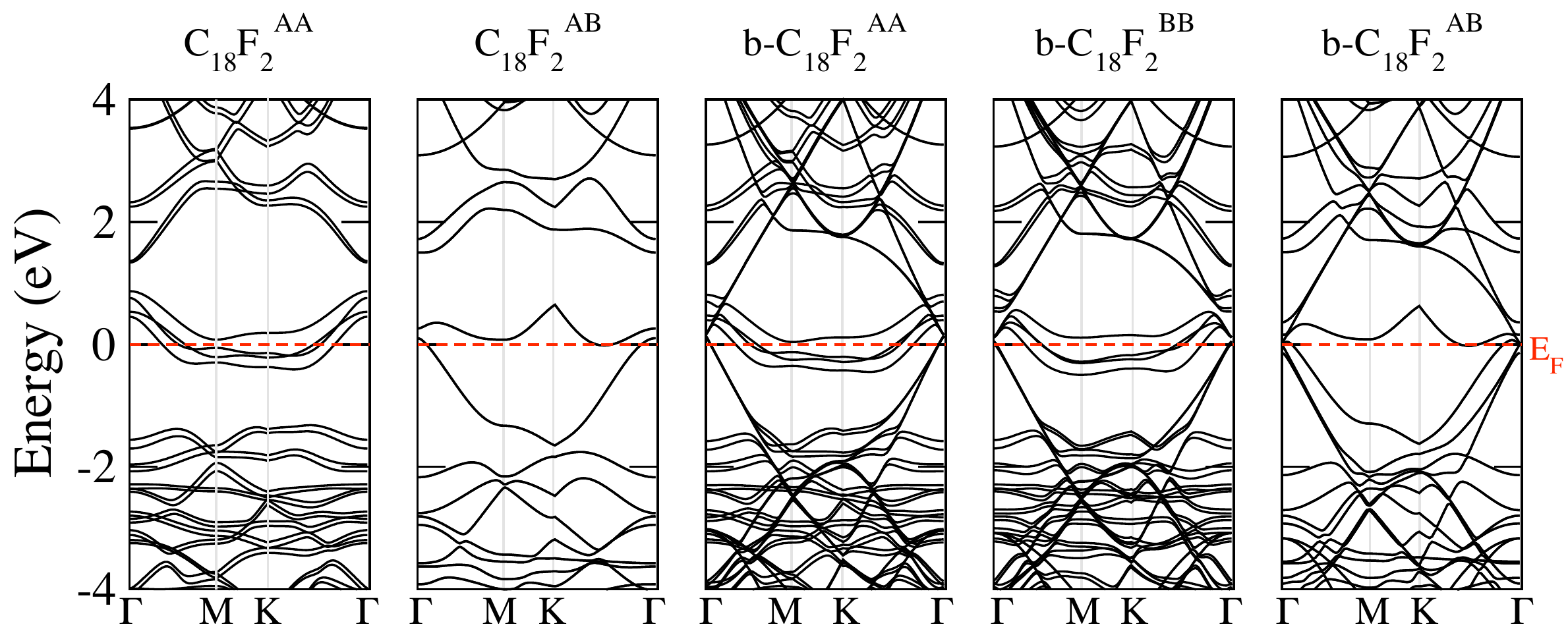}
  \caption{(Color online) Band structure of a) C$_{18}$F$_2^{AA}$, b) C$_{18}$F$_2^{AB}$, c) {\it b}-C$_{18}$F$_2^{AA}$, d)
  {\it b}-C$_{18}$F$_2^{BB}$, and e) {\it b}-C$_{18}$F$_2^{AB}$. Red dashed line indicates the Fermi level (E$_F$).  }
  \label{fig5_su}
 \end{figure}

In this supporting information we provide additionnal band structures
of fluorinated bi-layer graphene as a further illustration of the
electronic decoupling that occurs when the F is chemisorbed on one of
the two layers (Figs. \ref{fig1_su}, \ref{fig2_su}, \ref{fig3_su}, \ref{fig4_su}, and \ref{fig5_su}).

\end{document}